\documentclass{article}
\usepackage{spconf,amsmath,graphicx}
\usepackage{color}
\usepackage{url,hyperref,cleveref}
\usepackage{multicol,multirow}
\usepackage{diagbox} 
\usepackage{booktabs}
\usepackage{pifont}
\usepackage{makecell}
\usepackage{enumitem}
\usepackage{tablefootnote}
\usepackage[normalem]{ulem}

\title{Semi-supervised Learning for Code-Switching ASR with Large Language Model Filter}
\name{Yu Xi$^{1,2}$\thanks{This work was done during Yu Xi's internship at NVIDIA. Wen Ding$^\dagger$ is the corresponding author.}, Wen Ding$^{2{\dagger}}$, Kai Yu$^1$, Junjie Lai$^2$}
\address{$^1$Shanghai Jiao Tong University, Shanghai China\\ $^2$NVIDIA Corporation, Shanghai China \\
\url{{yuxi.cs, kai.yu}@sjtu.edu.cn}, \url{{wend, julienl}@nvidia.com}}
\begin{document}
%
\maketitle
\begin{abstract}
Code-switching (CS) phenomenon occurs when words or phrases from different languages are alternated in a single sentence. Due to data scarcity, building an effective CS Automatic Speech Recognition (ASR) system remains challenging. 
In this paper, we propose to enhance CS-ASR systems by utilizing rich unsupervised monolingual speech data within a semi-supervised learning framework, particularly when access to CS data is limited. To achieve this, we establish a general paradigm for applying noisy student training (NST) to the CS-ASR task. Specifically, we introduce the \textit{LLM-Filter}, which leverages well-designed prompt templates to activate the correction capability of large language models (LLMs) for monolingual data selection and pseudo-labels refinement during NST. Our experiments on the supervised ASRU-CS and unsupervised AISHELL-2 and LibriSpeech datasets show that our method not only achieves significant improvements over supervised and semi-supervised learning baselines for the CS task, but also attains better performance compared with the fully-supervised oracle upper-bound on the CS English part. Additionally, we further investigate the influence of accent on AESRC dataset and demonstrate that our method can get achieve additional benefits when the monolingual data contains relevant linguistic characteristic.

\end{abstract}
\begin{keywords}
Code-switching ASR, semi-supervised learning, noisy student training, large language model filter
\end{keywords}
\vspace{-10pt}
\section{Introduction}
\vspace{-5pt}
\label{sec:intro}

Semi-supervised learning (SSL), which leverages both supervised and unsupervised data, is a prominent deep learning approach extensively adopted in the field of Automatic Speech Recognition (ASR)~\cite{ssl_asr0,ssl_asr1,synnaeve2020endtoend}. Among various SSL techniques, Noisy Student Training (NST) has gained considerable attention due to its effectiveness to achieve state-of-the-art performances across multiple datasets~\cite{zhang2020pushing, park20d_interspeech}. Moreover, NST has proven beneficial in other aspects of ASR, such as domain adaptation~\cite{hwang2022large, intro-nst_filter-icassp2023} and low-resourced speech recognition~\cite{li-vu-2024-improving}. 
But for code-switching (CS) ASR, known for a scarcity of low-resource natural data as well as complex acoustic environments like accent shifting, there exist few research works under the SSL framework in recent years, especially the NST training strategy.

Over 60\% of people globally can speak more than one language and they often mix words or phrases from different languages within a single sentence~\cite{intro-60per-icassp2021-1,intro-60per-icassp2024-2}, allowing for natural code-switching. However, it is impractical to collect a large amount of transcribed CS data for ASR models since it requires hiring numerous transcribers with strong multilingual skills, resulting in a scarcity of natural CS data.   
Although advanced speech foundation models such as XLS-R~\cite{intro-xlsR-interspeech2021}, Whisper~\cite{intro-whisper-pmlr2023}, USM~\cite{intro-usm-arxiv2023}, SeamlessM4T~\cite{intro-sm4t-2023-arxiv,intro-sm4t2-2023-arxiv} and MMS~\cite{intro-MMS-meta-jmlr2024} pre-trained on large-scale multilingual speech data, can manage inter-sentential multilingual speech, but struggle with intra-sentential code-switching scenarios~\cite{intro-ftwhisper-for-csasr-icassp2024,intro-slm-struggle-arxiv2024}. 
Humans learn languages through monolingual materials but can seamlessly speak in such a CS fashion. Once proficient in multiple languages, they can naturally mix these languages within a sentence. Inspired by this, using various types of monolingual speech data to augment natural CS data and to build a good CS-ASR system that can integrate potential intra-sentential acoustic and linguistic information is more feasible.

\definecolor{dgrey}{RGB}{128,128,128}
\definecolor{lgrey}{RGB}{204,204,204}

\begin{figure*}[t]
\centerline{\includegraphics[width=16cm]{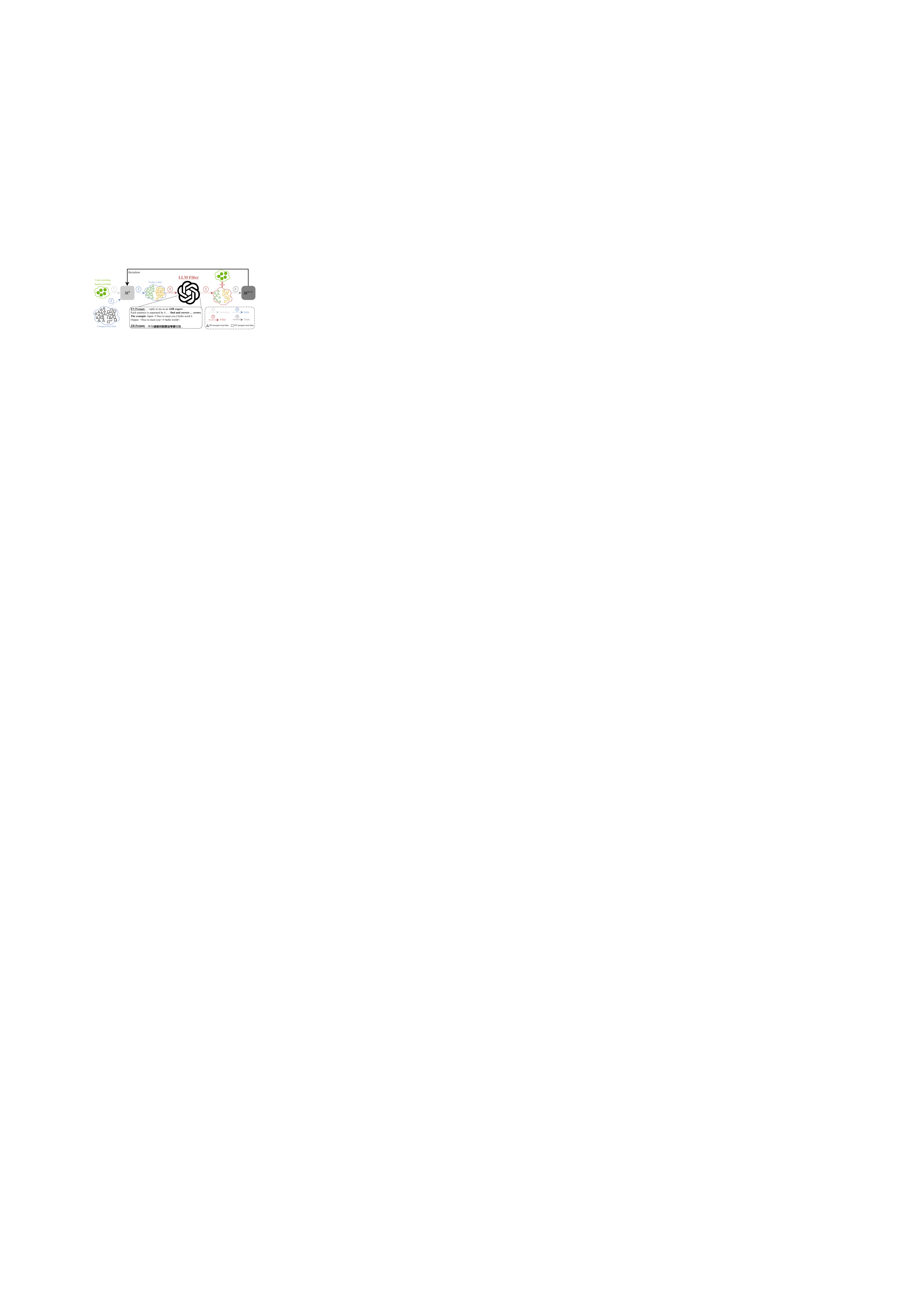}}
\caption{The overview of the entire NST with LLM-Filter pipeline involves two parts of data: a small supervised code-switching dataset and a large amount of unsupervised Mandarin and English monolingual data. Initially, we train the model ($M^{t}, t=0$) using only the CS data. Then, we use the CS data along with filtered high-quality pseudo-labeled EN and ZH monolingual data to train the \textcolor{dgrey}{\textbf{teacher model}} (dark grey). This is an iterative procedure, where the teacher model is assigned as the new \textcolor{lgrey}{\textbf{student model}} (light grey). Through repeated refinement and filtering of labels, we perform LLM-Filter boosted NST iterations until the student model converges. We also present an example prompt that is fed to the LLM. The ZH and EN prompts are essentially the same in meaning. For convenience, we only show the abbreviated version of the complete EN prompt.}
\label{fig:overview_figure}
\vspace{-1.4em}
\end{figure*}	

Currently, there are two main strategies to develop effective CS-ASR systems with monolingual data. The first type, based on language-aware mixture-of-experts (MOE)~\cite{intro-60per-icassp2021-1, lu20f_interspeech, tian22c_interspeech, zhang22x_interspeech}, is trained with multi-view or multi-task learning. These systems extract acoustic information using shared layers and process language-specific linguistic information using independent layers for each language. However, these approaches require more effort in data preparation and training strategy. Additionally, these models are designed only for the code-switching task, which leads to a lack of generalization.
The second type relies on monolingual data augmentation, which includes directly adding more high-resource supervised monolingual data~\cite{shan2019cslid, chuang2020trainingcs}, using online data translation between languages~\cite{liu2021csmono}, or intra-sentential CS speech simulation~\cite{du2021data, winata2019code}. Nevertheless, data mismatches could bring limited model improvement, low data efficiency, and high training costs. Thus, a more effective approach is to equip models with the flexibility to select data to improve their intrinsic code-switching performances. 
Applying data selection strategies for SSL is investigated in~\cite{intro-nst_lm_filter-interspeech2022,intro-nst_filter-icassp2023}. 
For instance, researchers in~\cite{intro-nst_filter-icassp2023} leverage the model difference between hypotheses from greedy decoding and N-Gram Language Model (LM) decoding to perform data selection on unsupervised out-of-domain speech data during NST iterations. 

In this paper, we pioneer the exploration of NST's potential to boost code-switching ASR performances. We follow a similar assumption and extend it to LLM by prompt engineering and in-context learning. LLMs like GPT-4o are used to construct a data filtering strategy to select unsupervised data with high-quality pseudo labels. 
There are several advantages of LLMs compared to traditional N-Gram based LMs: 1) Guided by scaling laws, LLMs are always pre-trained on a vast collection of text data, giving them strong abilities to model linguistic information. 
2) LLM is capable of processing multilingual texts; thus, only one LLM with different language prompts is needed instead of building numbers of monolingual LMs, simplifying the multilingual data selection pipeline.  
3) Due to their in-context learning ability and prompt engineering, we can design appropriate prompt templates to activate the hypothesis correction capabilities of LLMs, which is a more explicit way to interpret the errors from the ASR model compared to perplexities or confidence scores given by traditional LMs. 

In this work, \textit{LLM-Filter} is proposed to filter unsupervised multilingual audio data (in our experiments, English and Mandarin) for code-switching ASR. 
To our knowledge, this is the first research work to apply NST to code-switching ASR. Our contributions can be summarized as follows:

\begin{itemize}[itemsep=2pt,topsep=0pt,parsep=0pt]
    \item A general paradigm for NST in the CS-ASR scenario with supervised low-resource natural code-switching data and unsupervised monolingual data is introduced, which could alleviate the lack of CS data and is easy to extend to other languages. 
    \item Totally unsupervised data selection method \textit{LLM-Filter} is proposed during NST iterations, with multilingual prompts for one LLM alongside leveraging differences between greedy decoding and LLM corrected hypothesis as a threshold, which can utilize multiple monolingual data gradually and more efficiently. 
    \item Experiments on supervised CS English-Mandarin (EN-ZH) ASRU-CS, unsupervised AISHELL-2, and Librispeech datasets are conducted, which shows significant improvements over existing methods and even attains better performance on the CS English part comparable to the fully-supervised oracle upper-bound. 
    \item Switching unsupervised monolingual EN data to accented EN datasets AESRC demonstrates that our method can accomplish additional benefits when the monolingual data is more relevant to the CS domains.
    \item Codes and checkpoints open-sources in NeMo~\cite{kuchaiev2019nemo}. 
\end{itemize}

\vspace{-10pt}
\section{NST for Code-switching ASR}
\vspace{-5pt}
\label{sec:format}

NST is an extension of teacher-student learning. The student model is iteratively trained on supervised data and unsupervised data, the latter using pseudo-labels generated by the teacher model. Noise is introduced during the learning process to enhance the robustness of the student model. In our framework, the code-switching data serve as the supervised component, while the monolingual data are unsupervised using pseudo-labels, as shown in~\Cref{fig:overview_figure}. 

First, the seed model is trained on the supervised CS data (represented as circles) to \textbf{initialize} the NST iterations. Then, the seed model takes the teacher model role, conducting \textbf{inference} on the monolingual speech data (depicted as blank triangles and squares) to generate pseudo-labels for the unsupervised monolingual data. 
Unlike conventional NST approaches that directly combine supervised and unsupervised data to train the student model, our method incorporates a crucial \textbf{data filtering} step to assess the quality of pseudo-labels of the monolingual data. The details of our data selection process are presented in Section \ref{sec:LLM-filter}. 
Following the filtering stage, both the supervised CS data and filtered high-quality unsupervised monolingual data are fed into the student model. 
The student model will \textbf{iterate}, adopting the new teacher model role to infer the new pseudo-labels until convergence.

\vspace{-10pt}
\section{Large Language Model Filter}
\vspace{-6pt}
\label{sec:LLM-filter}
\subsection{Why do we use LLMs?}
LLMs are a series of game-changer works with remarkable capabilities that have re-formulated new research paradigms across various research fields, including ASR. With rich linguistic information learned from vast datasets, LLMs excel in tasks such as sentence refinement and grammatical corrections, which offers the potential for the correction of code-switching hypotheses. Particularly, they offer promising solutions for correcting code-switching hypotheses by addressing both spelling mistakes, thereby rectifying substitution errors and semantic inaccuracies to resolve deletion and insertion errors.
Several studies have enhanced ASR systems utilizing LLMs \cite{chen2023generative, pu2023multi,liu2024aligning,chenchen-asr-llm-nips2023,chuanyangwu-asr-llm-icassp2024}. Most of them concentrate on treating LLMs as conventional language models for re-scoring the N-Best hypothesis or fine-tuning LLMs to better integrate with front-end ASR models. This typically requires extensive computational resources. In contrast, our approach only relies on the 1-best hypothesis for error correction without additional training, rendering our LLMs effectively plug-and-play and independent of the underlying ASR models. 



\begin{table}[b]
  \centering
  \vspace{-12pt}
  \caption{The preliminary results to evaluate the effectiveness of LLM corrections on unsupervised ZH and EN datasets.}
  \vspace{5pt}
  \newcolumntype{S}{>{\small}c}
  \begin{resizebox}{1.0\columnwidth}{!}
  {
    \begin{tabular}{c| c | c | c c  c }
      \toprule
       \multirow{2}{*}{\textbf{\makecell{Unsup.\\Dataset}}} & \multirow{2}{*}{\textbf{\# Utts}} & \multicolumn{1}{c|}{\textbf{Greedy}} & \multicolumn{3}{c}{\textbf{LLM}} \\
      
       \cmidrule(lr){3-3} \cmidrule(lr){4-6}

      ~ & ~ &\textbf{\makecell{WER=0~(\%)}} & \textbf{\makecell{WER=0(\%)}}   & \textbf{\makecell{Not worse\\ hypo.(\%)}} & \textbf{\makecell{More accurate\\ hypo.(\%)}} \\
       
      \midrule
      \makecell{AISHELL-2} & 2000 & 20.5 & 26.6 & 83.0 & 25.2 \\
      \makecell{LS-360} & 1000 & 20.4 & 30.8 & 75.4 & 44.2  \\
      \makecell{LS-500} & 1000 & 13.7 & 22.5 & 82.8 & 56.0  \\
      \midrule
      ALL & 4000 & 18.8 & 26.6 & 81.1 & 37.6 \\
      \bottomrule
    \end{tabular}%
   }
   \end{resizebox}
  \label{table:preliminary_exp_to_valid_llm_performance}

\end{table}

Prompt engineering within LLMs activates task-specific capabilities through strategic instruction design, utilizing question-answer pairs in prompts to enhance understanding of user instructions.
Based on that, language-specific prompts are designed to enable LLMs to generate corrected hypotheses as pseudo labels for unsupervised monolingual datasets across different languages. 
Specifically, we utilize Mandarin prompts for Mandarin ASR outputs and English prompts for English outputs, with the potential to easily expand to other languages. 
Using a single LLM with multiple language-specific prompts streamlines the entire evaluation process, making data selection straightforward and effective.

\vspace{-10pt}
\subsection{Investigate the power of LLM correction}
\vspace{-5pt}
In this section, we detail the process of designing prompts and evaluating their efficacy in applying LLM-based corrections to ASR outputs. Initially, a batch of greedy decoding hypotheses generated by the ASR model is incorporated into the LLM prompt. Then, the entire prompt is fed into LLM to undertake necessary ASR error corrections accordingly. A typical prompt template is organized in the following format:
\textit{System instruction + Problem description + Examples.}

Here is an example of such a prompt in English, adapted similarly for use in Mandarin:

\textit{Please confirm my requirement and reply to me \textbf{as an ASR expert}. I will provide a batch of decoding results of an ASR model. Each sentence is separated by \#, and you will assist me to \textbf{find and correct possible substitution, insertion, and deletion errors, and output the corrected result}. The final output format is $<$corrected result$>$. \textbf{The example is as follows}: Input: \#Nice to meat you\#hello word\#. Output: $<$Nice to meet you$>$\#$<$hello world$>$. After each time I input ASR hypotheses, please return the result directly without the inference process.}


To validate the effectiveness of LLMs-based correction, we randomly selected 2,000 monolingual Mandarin and English audio samples. The code-switching ASR base model generated greedy decoding hypotheses for each sample. We calculated the Word Error Rate~(WER) between these hypotheses and their ground truths and present the preliminary LLM correction results in ~\Cref{table:preliminary_exp_to_valid_llm_performance}. 
The ratios of total correct audios (WER=0) are illustrated, and the percentages of LLM improvement of hypotheses are evaluated, including cases where the LLM-corrected output is not worse or more accurate than the previous greedy decoding results.

The results indicate that approximately 26.6\% of hypotheses can be corrected by the LLM even without acoustic information, and around 37.6\% of them are partially corrected compared to the greedy results. Overall, around 81.1\% of the hypotheses are either preserved or enhanced in accuracy across various language datasets. These results underscore the significant enhancement in the quality of pseudo labels through LLM corrections, demonstrating the model's potential in refining ASR outputs.


\vspace{-10pt}
\subsection{LLM-Filter to select high-quality pseudo labels}
\vspace{-5pt}

\begin{figure}[t]
  \centering
    \includegraphics[width=\linewidth]{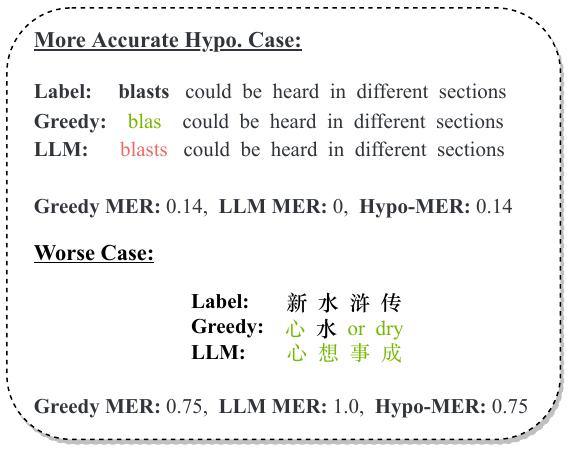}
    \vspace{-2.0em}
    \caption{Two examples to illustrate the ground-truth (Label), greedy decoding result (Greedy) and LLM corrected hypothesis (LLM). The MERs are calculated respectively. }
    \label{fig:llm_correction_examples}
    \vspace{-1.5em}
\end{figure}

Although there are significant quality improvements in pseudo labels through LLM corrections, leveraging these corrected outputs directly for NST still presents performance challenges. 
Upon analyzing a substantial volume of outputs, we summarized two patterns as shown in~\Cref{fig:llm_correction_examples}. 
In the first case, a single word error (`blas' $\rightarrow$ `blasts') is accurately correctly by the LLM, thereby perfecting the whole sentence.

Conversely, in cases where only one word is recognized correctly, LLM fails to improve the recognition results. We observed that LLM corrections are more effective when the initial greedy output is nearly accurate; however, they become less potent as the error rate of the greedy result increases. This diminished efficacy is likely due to the faulty linguistic context provided by erroneous greedy hypotheses, which misguide the LLM's predictions.
Building on insights from~\cite{intro-nst_filter-icassp2023}, we adopt the assumption that if a Large Language Model assesses a decoding hypothesis as accurate enough to require no further modifications, it is likely to constitute a high-quality pseudo label suitable for SSL training.
We assess the quality of pseudo labels by comparing the discrepancies between greedy and LLM-corrected hypotheses, utilizing the Character Error Rate (CER) for Mandarin and Word Error Rate (WER) for English. These metrics establish a threshold for data selection, denoted as \textbf{Hypo-CER/Hypo-WER}. For uniformity across languages, we employ the Mixed Error Rate (MER), referred to as \textbf{Hypo-MER}.

\vspace{-7pt}
\section{Experiments}
\vspace{-12pt}
\label{sec:typestyle}
\subsection{Datasets}
\vspace{-5pt}
We use three types of training datasets in our study:

\begin{itemize}[itemsep=2pt,topsep=0pt,parsep=0pt]
    \item \textbf{Code-switching data}. The ASRU 2019 Mandarin-English Code-Switching Challenge dataset~\cite{shi2020asru} (referred to as ASRU-CS or ASRU) serves as our supervised data. This dataset consists of around 200 hours of natural intra-sentential code-switching (CS) audios spoken by native Mandarin speakers in China. Notably, we only utilize the CS portion of the dataset, excluding the 500-hour Mandarin-only part.
    \item \textbf{Monolingual data}. AISHELL-2~\cite{du2018aishell}, referred to as ASL-2, is an open-source Mandarin corpus with 1000 hours of data. LibriSpeech~\cite{dan2015librispeech} (LS) consists of 960 hours of English monolingual speech, including two clean subsets: LibriSpeech-clean-100 and LibriSpeech-clean-360, and one noisier subset: LibriSpeech-other-500. We denote LibriSpeech-clean-100 as LS100 and use LS860 or LS960 to refer to different combinations of these subsets.
    \item \textbf{Accented English data}. The Accented English Speech Recognition Challenge dataset~\cite{aesrc_dataset_icassp2021} (AESRC) is a 160-hour speech corpus featuring accented English from 8 countries, including 20 hours of English with a Chinese accent. This dataset is regarded as an extra unsupervised English dataset for filtering due to its relevance to the ASRU-CS corpus we employed.
\end{itemize}

Evaluations are performed on CS data, including the dev and test sets of ASRU.
To further demonstrate the effectiveness of monolingual data, the results of dev and test sets of ASL-2, test-clean, and test-other sets of LS are also reported.

\begin{table*}[t]
  \centering
   \caption{The baselines results including supervised training with ASRU CS data and ASRU CS plus LS100 data, standard NST with unsupervised ASL-2 and LS860, and the upper-bound with all supervised data. We \textbf{bold} the supervised oracle results, and \uline{\textbf{bold and underline}} the best results of the proposed method. The performances of 3 iterations of NST with LLM-Filter are included. Besides the MERs on the test sets of ASRU, the results on monolingual test sets are reported as well. }
   \vspace{5pt}
  \newcolumntype{S}{>{\small}c}
  \begin{resizebox}{2.0\columnwidth}{!}
  {
    \begin{tabular}{c | c| c  c | c | c  c c c c c| c c | c c}
      \toprule
       ~ & ~ &   \multicolumn{2}{c|}{\textbf{Data}} &\multirow{3}{*}{\textbf{\makecell{LLM\\Filter}}} & \multicolumn{6}{c|}{\textbf{ASRU-CS(ASRU)}} & \multicolumn{2}{c|}{\textbf{AISHELL-2(ASL-2)}}& \multicolumn{2}{c}{\textbf{LibriSpeech(LS)}} \\
      
       \cmidrule(lr){3-4}\cmidrule(lr){6-11} \cmidrule(lr){12-13} \cmidrule(lr){14-15}

      \textbf{ID}  & \textbf{Model} &\multirow{2}{*}{\textbf{\makecell{Supervised\\ labels}}} &\multirow{2}{*}{\textbf{\makecell{Unsupervised \\ pseudo labels}}} & ~ & \multicolumn{3}{c}{dev} & \multicolumn{3}{|c|}{test} &  \multicolumn{1}{c}{\quad dev}  & \multicolumn{1}{c|}{\quad test}  & \multicolumn{1}{c}{test-clean} & test-other \\
       
    \cmidrule(lr){6-8} \cmidrule(lr){9-11} \cmidrule(lr){12-13} \cmidrule(lr){14-15}
      ~& ~ & ~&~ & ~& MER & CER & \multicolumn{1}{c|}{WER} & MER & CER & WER & \multicolumn{2}{c|}{CER}  & \multicolumn{2}{c}{WER} \\
     
      \midrule
     A &   Sup. Base & ASRU & -  & \ding{56} & 16.20 & 13.67 & 39.33 & 15.57 & 13.08 & 38.50  & \quad 26.57 & \quad 27.06 & 96.75  & 98.86  \\ 
       
     B &  NST Seed Base & ASRU + LS100 & - & \ding{56} & 15.01 & 13.16 & 32.75 & 14.27 & 12.44 & 31.52  & \quad 25.47 & \quad 26.11 & 9.19  & 24.25 \\
     C &  Semi-sup. Base & ASRU + LS100 & ASL-2 + LS860 & \ding{56} & 13.91 & 11.94 & 32.59 & 13.29 & 11.4 & 31.13 & \quad 19.96 & \quad 20.89 & 7.27 & 17.83 \\
       \midrule
     D &  Oracle Base  & \makecell{ASRU + LS960\\ + ASL-2}  & - & \ding{56} & \textbf{10.99} & \textbf{8.82} & \textbf{30.92} & \textbf{10.45} & \textbf{8.42} & \textbf{29.37} & \quad \textbf{7.77} & \quad \textbf{8.12} & \textbf{4.50} & \textbf{10.56}  \\
      
      \midrule
      
    E&  \multirow{3}*{NST}&\multirow{3}*{ASRU + LS100} & \multirow{3}*{ASL-2 + LS860} & \textbf{Iter 1} &13.33 & 11.44 & 31.2 & 12.64 & 10.79 & 30.14 & \quad 18.12 & \quad 18.74 & 7.49 & 18.11\\
    F&  ~& ~&~ &  \textbf{Iter 2} & 12.44 & 10.62 & 29.72 & 11.79 & 10.04 & 28.25 & \quad \uline{\textbf{16.09}} & \quad 16.61 & 4.50 & \uline{\textbf{10.84}}\\
     G & & ~&~ & \textbf{Iter 3} & \uline{\textbf{12.11}} & \uline{\textbf{10.32}} & \uline{\textbf{28.98}} & \uline{\textbf{11.59}} & \uline{\textbf{9.84}} & \uline{\textbf{28.16}} & \quad 16.16 & \quad \uline{\textbf{16.46}} & \uline{\textbf{4.47}} & 11.08 \\
      \bottomrule
    \end{tabular}%
   }
   \end{resizebox}

  \label{table:nst_results_on_asl2_and_ls}
\vspace{-0.8em}
\end{table*}

\vspace{-10pt}
\subsection{Experimental setup}
\vspace{-5pt}
We choose RNN-T as our model structure and use the parameter configuration of Fast-Conformer~\cite{rekesh2023fast} in NeMo~\cite{kuchaiev2019nemo} for all ASR experiments. The model consists of 17 layers with a 512-dimension Fast-Conformer encoder and 640-dimensional 2 layers of LSTM decoder. Additionally, the hidden dimension of the joint network is 640. The total number of parameters is about 120 million. The model structures of the teacher and student are the same. 

The audio features consist of 80-dim log Mel-filter bank coefficients (FBank) extracted using a 25ms window with a 10ms window hop. SpecAugment~\cite{park2019specaugment} is applied during training, with a maximum frequency mask range of 27 and a maximum time mask range of 50ms. The concatenated tokenizer~\cite{dhawan-etal-2023-unified} is constructed with 3981 Chinese characters and 1024 English BPEs, totaling 5005 tokens.

We use the AdamW~\cite{adam-iclr2015,adamw-Ilya-frank-iclr2019} optimizer with a maximum learning rate (lr) of 7.5e-4. The Cosine Annealing~\cite{cosineannealing-Ilya-frank-iclr2017} lr scheduler is applied, with warm-up steps set to 10,000 across all experiments. The mini-batch size is 16, and the best five checkpoints are averaged. For efficiency, if there is no improvement in several epochs, the training will be terminated, and the maximum number of epochs is set to 100. We train all models using 8 NVIDIA V100 GPUs in parallel.

We choose the powerful LLM, GPT-4o, developed by OpenAI, as our LLM-Filter to refine the quality of pseudo labels. Each interaction with the GPT-4o API involves a batch of 40 greedy hypotheses, up to a maximum of three attempts. If it fails more than three times, we drop the batch. Hypo-MER is set to 0.1 to filter suitable pseudo labels for NST iterations. In order to balance the amount of monolingual datasets, we allocate equal durations of pseudo-labeled Chinese~(ZH) and English~(EN) data for training our NST model after applying the LLM-Filter.

\vspace{-10pt}
\subsection{Baselines}
\vspace{-5pt}

In this section, we outline the baselines as detailed in the upper section of~\Cref{table:nst_results_on_asl2_and_ls}. All hyper-parameters, with the exception of the training datasets, are kept consistent across these baselines to ensure comparability.

\textbf{Supervised baseline (Sup. Base)}. This baseline utilizes the 200-hour code-switching (CS) ASRU dataset for training the ASR model. It serves not only as the initial point of comparison for training on CS supervised data but also as the foundational baseline for our subsequent NST LLM-Filter experiments. However, considering that the WERs for the English monolingual dataset, LibriSpeech, approach nearly 100\% (with most misrecognized as Mandarin characters), it is impractical to use such inaccurate hypotheses for data filtering. Therefore, we propose an alternate NST seed baseline.

\textbf{NST seed baseline}. To this model, we add an additional 100 hours of supervised English data from LibriSpeech LS100 data to the 200-hour ASRU dataset. This modification can mitigate the imbalance between Mandarin (ZH) and English (EN) in the CS-ASR model. We utilize this augmented model as the seed for subsequent NST iterations. Compared to the Sup. Base model, this baseline shows improved performance on both the English segments of the ASRU dataset and the LibriSpeech test sets.

\textbf{Standard NST as Semi-supervised baseline (Semi-sup. Base)}. This baseline represents the standard SSL approach using NST. Pseudo labels are generated by performing inference on AISHELL-2 and the remaining 860 hours of LibriSpeech (LS860) data (comprising LS-clean-360 and LS-other-500) using the NST seed baseline. A student model is then trained from scratch using both supervised ASRU and LS100 datasets, as well as the pseudo-labeled AISHELL-2 and LS860 datasets. In summary, there are 300 hours of supervised data and 1860 hours of unsupervised data. 

\textbf{Oracle baseline (Oracle Base)}. As a fully labeled upper-bound baseline, this model is trained on a total of 2160 hours of supervised data, which includes 200 hours from the CS ASRU dataset, 1000 hours of ZH data from AISHELL-2, and 960 hours of EN data from LibriSpeech. We use this baseline to gauge the upper limits of performance and identify potential areas for improvement in our proposed SSL models.

\vspace{-8pt}
\section{Results Analysis}
\vspace{-4pt}
\label{sec:results}

\vspace{-3pt}
\subsection{Performances on NST with LLM-Filter }
\vspace{-5pt}
Firstly, we investigate the impact of incorporating monolingual unsupervised data in the CS task. Compared to Model B, the supervised CS baseline, the addition of unsupervised ZH from AISHELL-2 and unsupervised EN from LibriSpeech proves beneficial. This enhancement allows the standard NST baseline, Model C, to achieve gains of 7.3\% and 6.7\% over Model B (B\;vs.\;C). This improvement is consistently observed across the test sets of both monolingual datasets. However, there remains a substantial performance gap relative to the fully supervised baseline, Model D. This gap arises from the noise introduced by low-quality pseudo labels, which hinders optimal model convergence.

Next, we evaluate the efficacy of the proposed LLM-Filter, as detailed in the last three rows of~\Cref{table:nst_results_on_asl2_and_ls}. As the number of LLM-Filter rounds increases, the performance on the ASRU test set improves (E\;vs.\;F vs.\;G). Significant improvements are also observed in both monolingual Mandarin (ZH) and English (EN) datasets, indicating that the LLM-Filter effectively enhances the quality of pseudo labels. 
The performance gain decreases unsurprisingly because the quality of the pseudo labels is already good enough so that the model could not learn much more from additional iterations. Furthermore, compared to the standard NST without data filtering (C\;vs.\;G), significant improvements of 12.6\% and 12.9\% in MER can be achieved in the ASRU dev and test datasets, respectively. Notably, due to the high computational and LLM inference cost, we only ran three iterations. We believe that although the gain has decreased, with more iterations, the gap between the NST model with LLM-Filter and the oracle baseline can be further narrowed.

\begin{figure}[ht]
  \centering
    \includegraphics[width=\linewidth]{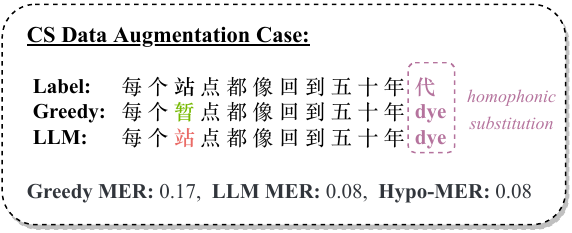}
    \vspace{-2.0em}
    \caption{A homophonic substitution case in ZH training dataset, which can be regarded as an automatic data augmentation strategy to boost CS ASR training.}
    \label{fig:cs_aug_example}
    \vspace{-2.0em}
\end{figure}

When comparing the oracle model D to our model G with LLM-Filter, it is surprising that our NST iterative model outperforms the oracle baseline on the EN part of the CS ASRU and on the monolingual EN LibriSpeech test datasets. This unexpected outcome can be attributed to the preservation of high-quality homophonic substitution pseudo labels by the LLM-Filter. An example is presented in~\Cref{fig:cs_aug_example}. 

Our proposed NST method with LLM-Filter generates high-quality intra-sentential CS data, effectively serving as a CS data augmentation strategy. This method alleviates the data scarcity problem and acts as a potential supplement that traditional supervised monolingual datasets cannot provide.

\vspace{-0.8em}

\vspace{-5pt}
\subsection{The influence of accent dataset AESRC}
\vspace{-5pt}

A comparison analysis of the results between with (w/) and without (w/o) the accent dataset AESRC is presented in~\Cref{table:add_accent_data_aesrc}. Following the second iteration on ASL-2 and LS, we observed a significant decrease in the WER for LS, while the WER of the ASRU EN part decreased more gradually. We attribute the inconsistency in the EN part primarily to the mismatch between the non-accented LS dataset and the Chinese-accent ASRU dataset. To address this, we then replace part of EN unsupervised data with the accented AESRC dataset and adopt the same LLM-Filter strategy. Given the shorter total duration of speech in AESRC, we balanced the dataset volumes by selectively using a subset of LS for training.

The results show that, although the overall MER slightly increases, the inclusion of the accented dataset substantially enhanced the performance of the English component in the CS model (28.16\%\;vs.\;27.51\%). It indicates that LLM-Filter NST is more effective for CS-ASR when the monolingual dataset closely aligns with the linguistic characteristics of the target dataset. In addition, the overall MER increase is reasonable, considering that the WER for the AESRC training dataset is 21.61\%, significantly higher than LS's 4.04\% WER (even higher than the 13.33\% WER from the initial LS round)\footnote{For simplicity, we don't report this part of results in the paper.}. 
The higher WER introduces more noise into the NST process, impacting the training for the Mandarin component. However, we believe that conducting more NST iterations on AESRC would likely yield better overall results, allowing the model to better accommodate the accented English data.

      

       
      


\begin{table}[h]
  \centering
  \vspace{-1.0em}
   \caption{Comparisons of switching EN LS860 unsupervised data to AESRC on CS ASRU test dataset. }
    \vspace{5pt}
  \newcolumntype{S}{>{\small}c}
  \begin{resizebox}{1.0\columnwidth}{!}
  {
    \begin{tabular}{c| c  c | c  c c}
      \toprule
        \multirow{2}{*}{\textbf{Model}} &\multirow{2}{*}{\textbf{\makecell{Sup.\\ labels}}} &\multirow{2}{*}{\textbf{\makecell{Unsup. \\ pseudo labels}}} & \multirow{2}{*}{\textbf{MER}} & \multirow{2}{*}{\textbf{CER}} & \multirow{2}{*}{\textbf{WER}} \\
     ~& ~ & ~& ~ & ~ & ~ \\
      
      \midrule
      \makecell{NST Iter 2 \\ (Seed for Iter 3)} & \makecell{ASRU\\ + LS100} & \makecell{ASL-2\\+ LS860} &  11.79 & 10.04 & 28.25 \\
      \midrule\midrule

      NST Iter3 & \makecell{ASRU\\ + LS100} & \makecell{ASL-2\\+ LS860} &  \textbf{11.59} & \textbf{9.84} & 28.16 \\
      \midrule
      
      \makecell{NST Iter3} & \makecell{ASRU\\ + LS100} & \makecell{ASL-2 + LS860 \\ + AESRC} & 11.80 & 10.14 & \uline{\textbf{27.51}} \\
      \bottomrule
    \end{tabular}%
   }
   \end{resizebox}
  \label{table:add_accent_data_aesrc}
\vspace{-1.4em}
\end{table}

\vspace{-6pt}
\subsection{The assessment of data filtering }
\vspace{-5pt}
The duration of filtered unsupervised data in each NST iteration and the ratios of filtered hours to total available hours are presented in Table~\ref{table:filterd_data_analysis}. The CER for greedy decoding results (Greedy CER) and LLM-corrected results (LLM CER) are also calculated. 
From Iter1 to Iter2, the filtered hours approximately double, illustrating the gradual and effective selection of data by our proposed method. Concurrently, the CERs for these data decrease, which enhances the model's ability to learn from more accurate pseudo labels. 
Despite an apparent upper-bound on the data ratio, as the slight increase in filtered ratio stagnates from Iter2 to Iter3, the quality of the pseudo labels continues to enhance. 
In Iter3, only 463.93 hours, about half of the unsupervised data are used for training compared to the semi-supervised baseline Model C. 
This observation indicates the cost-effectiveness of our proposed data filter strategy in utilizing unsupervised data and significantly accelerating the training process of SSL by focusing on higher-quality inputs.


\begin{table}[ht]
  \centering
  \vspace{-12pt}
  \caption{The quantity of pseudo labels for training set of AISHELL-2 before NST and in NST iterations. The threshold of LLM filter,Hypo-MER, is 0.1 by default.}
  \vspace{5pt}
  \newcolumntype{S}{>{\small}c}
  \begin{resizebox}{1.0\columnwidth}{!}
  {
    \begin{tabular}{c c | c c c c c c c c }
      \toprule
      
       \multirow{2}*{\textbf{Model}}  & \multirow{2}*{\textbf{\makecell{LLM\\Filter}}} &  \multirow{2}*{\textbf{\makecell{Total\\ hours}}} &  \multirow{2}*{\textbf{\makecell{Filtered\\ hours}}} & \multirow{2}*{\textbf{\makecell{Filtered\\ratio}}}&  \multirow{2}*{\textbf{\makecell{Greedy \\ CER}}} & \multirow{2}*{\textbf{\makecell{LLM Filtered\\ CER}}}\\
       ~&~&~&~&~ \\
      \midrule
      NST Seed Base & \ding{56} & 1000 & 1000 & 1 & 32.08 & 32.08 \\
      \midrule
      NST Iter1 & \ding{51} & 686.93 & 266.82 & 0.39 & 31.74 & 21.37 \\
      NST Iter2 & \ding{51} & 833.97 & 526.17 & 0.63 & 21.94 & 15.52 \\
      NST Iter3 & \ding{51} & 721.58 & 463.93 & 0.64 & 18.31 & 13.39 \\
      \bottomrule
    \end{tabular}%
   }
   \end{resizebox}
  \label{table:filterd_data_analysis}
\vspace{-7pt}
\end{table}

\vspace{-12pt}
\section{Conclusions}
\vspace{-5pt}
\label{sec:conclusions}
In this paper, an efficient and straightforward data selection strategy \textit{LLM-Filter} is proposed when applying the SSL techniques to the CS-ASR scenario. Experiments on the supervised CS ASRU dataset and unsupervised monolingual ASL-2 and LS are conducted. 12.6\% and 12.9\% significant MER reductions can be achieved in the test sets of ASRU while only half of the unsupervised data are utilized for training compared to the standard NST without data filtering. Performances can be further improved when adding accented EN dataset AESRC, which has more relevant domains with the supervised data. Future work will be to investigate LLM-Filter on the pretrained speech foundation models. 


\bibliographystyle{IEEEbib}
\bibliography{citations/refs}

\end{document}